\documentstyle[psfig]{mn}

\def\refnew#1{(\ref{#1})}
\def\etal{et al.\,}
\def\be{\begin{equation}}
\def\ee{\end{equation}}

\def\K{\, \rm K}
\def\s{\, \rm s}

\begin{document} 

\title{Combination Frequencies in the Fourier Spectra of  White Dwarfs}
\author[Yanqin Wu]
	{Yanqin Wu \\Astronomy Unit, School of Mathematical Sciences,
	Queen Mary and Westfield College, Mile End Road, London E1
	4NS, UK\\{email address:wu@cita.utoronto.ca}}


\maketitle

\begin{abstract}
Combination frequencies are observed in the Fourier spectra of
pulsating DA and DB white dwarfs. They appear at sums and differences
of frequencies associated with the stellar gravity-modes.  Brickhill
(1992) proposed that the combination frequencies result from mixing of
the eigenmode signals as the surface convection zone varying in depth
when undergoing pulsation. This depth changes cause time-dependent
thermal impedance, which mix different harmonic frequencies in the
light curve.

Following Brickhill's proposal, we developed {\it analytical}
expressions to describe the amplitudes and phases of these combination
frequencies.  The parameters that appear in these expressions are: the
depth of the stellar convection zone when at rest, the sensitivity of
this depth towards changes in stellar effective temperature, the
inclination angle of the stellar pulsation axis with respect to the
line of sight, and lastly, the spherical degrees of the eigenmodes
involved in the mixing. Adopting reasonable values for these
parameters, we apply our expressions to a DA and a DB variable white
dwarf. We find reasonable agreement between theory and observation,
though some discrepancies remain unexplained. We show that it is
possible to identify the spherical degrees of the pulsation modes
using the combination frequencies.

\end{abstract}


\section{Introduction
\label{sec:LtCv-observ}}

Long sequences of almost uninterrupted light curves, obtained during
the Whole Earth Telescope (WET) campaign on the helium variable white
dwarf GD358 (Winget \etal 1994, for a more recent analysis, see Vuille
\etal 2000), disclosed not only the presence of a large number of
stellar pulsational modes in the Fourier spectra, but also the
presence of `combination frequencies', signals that lie at the sum or
difference frequencies of the stellar eigen-modes (see Fig. 7 in the
first paper).

Combination frequencies have been observed in other pulsating white
dwarfs with either hydrogen or helium atmospheres, e.g., ZZ Psc (aka
G29-38, McGraw 1976; Kleinman \etal 1998), GD154 \cite{robinson78},
BPM31594 (McGraw 1976; O'Donoghue, Warner \& Cropper 1992), G117-B15A
\cite{kepler82} and GD165 \cite{bergeron93}.  Indeed, every variable
hydrogen white dwarf (class name ZZ Ceti, or DAV) that has been
observed with sufficiently high signal-to-noise ratio exhibits
combination frequencies (Brassard, Fontaine \& Wesemael 1995). The
same likely holds for helium variables (DBV).

Combination frequencies are thought to result from nonlinear mixing of
sinusoidal signals that are associated with the eigenmodes (named the
`principal modes' in this article). This conclusion is based on the
following arguments: combination frequencies are too numerous to be
eigenmodes themselves (Winget \etal 1994); amplitudes of the
combination frequencies have been shown to correlate with those of
their principal modes (for an early review, see McGraw 1978);
combination frequencies tend to have more complicated fine structure
than their principal modes, which can be explained naturally by a
linear superposition of the principal modes' rotationally split
multiplets \cite{winget94}.

Brickhill \shortcite{brick92} showed that nonlinear mixing arises
naturally in the context of his theory of convective driving
(Brickhill 1983, 1990, 1991a, 1991b). He realized that the convective
turn-over time scale in DA and DB variable white dwarfs is much
shorter than the pulsation period. Thus one can safely assume that the
surface convective region adjusts instantaneously during
pulsation. Brickhill found that under this assumption, the
photospheric flux variation is delayed and reduced relative to that
entering the bottom of the convection zone, by an amount depending on
the depth of the convection zone. Instantaneous adjustment of the
convection zone also implies that the extent of the convection zone
varies during the pulsation cycle, thus leading to variations in the
reduction and delay of the flux variation. This distorts the shape of
the light curve at the photosphere, and brings about the combination
frequencies in the Fourier power spectrum. Using a numerical analysis,
Brickhill found that for reasonable amplitudes of the principal modes,
he could reproduce the observed amplitudes of the combination
frequencies.  Note that in this theory, the combination frequencies
reflect distortion of the light curve by the nonlinear medium; they
are not associated with physical displacements and
velocities.\footnote{Fast convection enforces uniform movement
throughout the convective region \cite{paperIII}. There is no
distortion to the velocity signal.} This is indeed confirmed
observationally (van Kerkwijk, Clemens \& Wu 1999).

In this paper, we use a perturbative analysis to derive analytical
formulae for the strength and phase of the combination
frequencies. The advantage of this analysis over Brickhill's numerical
approach is that the dependence on stellar properties becomes
explicit. We find that two parameters, namely, the depth of the
surface convection zone when the star is at rest, and the sensitivity
of this depth towards changes in stellar effective temperature,
determine the efficiency of the mixing process. We also show that two
geometric factors, the spherical degrees of the principal modes, and
the inclination angle of the stellar pulsation axis, enter the
analytical expressions.  We compare our formulae with data on GD358 (a
DBV) and G29-38 (a DAV), adopting appropriate values for the above
stellar parameters. Despite imperfect agreement, we show that it is
possible to infer the spherical degree for the principal modes. In \S
\ref{subsec:othertype}, we briefly discuss the prospects of explaining
the combination frequencies in other types of variable stars.

\section{Perturbation Analysis}
\label{sec:LtCv-theory}

\subsection{Origin
\label{sec:LtCv-origin}}

In this section, we demonstrate that the surface convection zone can
nonlinearly mix sinusoidal flux variations to produce combination
frequencies.

\begin{figure}
\centerline{\psfig{figure=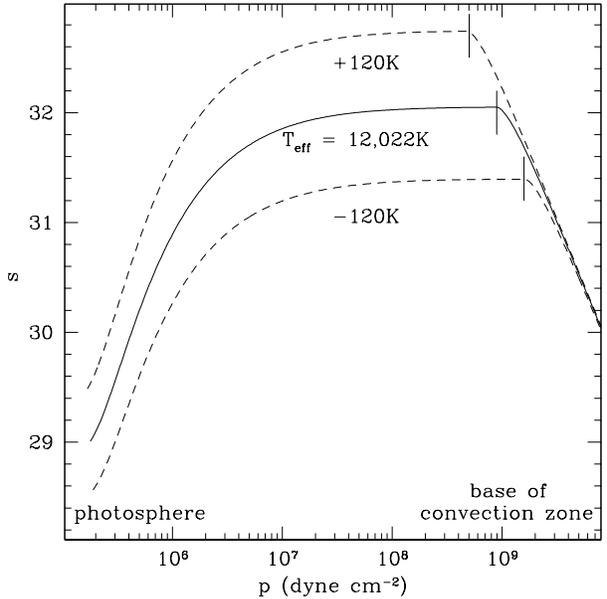,width=8.5cm}}
\caption[]{Entropy  as a function of  pressure 
in the upper atmosphere of three hydrogen white dwarf models with
similar effective temperatures \cite{paperII}.  The bottom of the
convection zone (marked with short vertical bars) shifts by one
pressure scale-height when the surface flux varies by $\sim
4\%$. There is a fast convergence in entropy towards the radiative
interior. The entropy here is dimensionless and is related to the
physical one by a factor of $k_B/m_p$, with $k_B$ the Boltzmann
constant, and $m_p$ the proton mass. }
\label{fig:LtCv-sprofile}
\end{figure}

We adopt the following three simplifying assumptions.  Firstly, we
assume neighbouring angular directions do not affect each
other. Observed pulsations in white dwarfs are associated with
eigenmodes of low spherical degree ($\ell = 1$ or $2$). Horizontal
variations in these modes occur over a scale of order the stellar
radius. One can safely ignore the effect of horizontal heat
transfer. This assumption allows us to study one angular direction and
later generalize the results to other directions without modification.

Secondly, we restrict our analysis to the surface convective region
and assume that the radiative interior caused little nonlinearity in
the pulsation signals. In the deep radiative interior where the
pulsation is adiabatic, the relative importance of the nonlinearity is
measured by the local pressure perturbation, $(\delta p/p)$, which is
smaller than a few percent for all observed modes. In the radiative
region immediately below the convection zone, the situation is less
clear. However, it seems reasonable to assume that the nonlinearity in
this region is weak compared to that arising from the convective
region as the reaction time of the former region (local thermal time)
is much longer than that in the latter (see later). Further study is
necessary to confirm or to refute this assumption.
This assumption allows us to take the flux perturbation incident upon
the bottom of the convection zone, $(\delta F/F)_b$, to be
sinusoidal. This also allows us to ignore entropy variations in the
radiative region (unless the material becomes convective).

Thirdly, we adopt equilibrium models that are adjacent in effective
temperature to quantify the time-dependent nature of the convective
region. The three models in Figure \ref{fig:LtCv-sprofile} resemble a
radial column of gas at flux maximum, passing through the rest point,
and at flux minimum, respectively. This simplification is possible
because the thin convection zone in a pulsating white dwarf has eddy
turn-over times much shorter than the pulsation period and can react
instantaneously to pulsation. The entropy of the convection zone
varies in phase with the surface flux perturbation (this is an
important feature for driving the gravity-modes, see Brickhill 1983,
1990; Goldreich \& Wu 1999, hereafter Paper I), the size of the
convection zone adjusts accordingly. In this analysis, we consider
changes in the depth of the convection zone caused by entropy
variations only. Other effects, e.g., convective overshoot and shear
turbulence, are not taken into account.

Under these assumptions, we derive the flux perturbation emerging at
the photosphere, $(\delta F/F)_{\rm ph}$, for given $(\delta
F/F)_b$. These two flux perturbations are related through energy
conservation,
\begin{equation}
\left({{\delta F}\over F}\right)_{\rm ph}   =
\left({{\delta F}\over F}\right)_{b} - {1\over F} {{d\Delta Q}\over{dt}},
\label{eq:firstlayprelim}
\end{equation}
where $\Delta Q$ is the amount of excess heat (compared to that at
equilibrium) stored in the convective region, and $F$ is the stellar
flux at equilibrium. 

Denoting the depth of the convection zone at time $t$ as $z_b(t)$, we
find
\begin{eqnarray}
\Delta Q & = & \int_{0}^{z_b (t)} dz \, \rho T {{k_B}\over{m_p}} (s - s_0) 
\nonumber \\
& \approx & F \tau_{b_0} \Delta s + {{k_B}\over{m_p}} \int_{z_{b_0}}^{z_b (t)}
dz \, \rho T  (s - s_0),
\label{eq:LtCv-dQall} 
\end{eqnarray}
where the second term on the right hand side is small compared to the
first term. Here, all quantities marked with subscript $0$ are
equilibrium quantities, and the thermal time constant $\tau_b$ at
depth $z_b$ is defined as (see Paper I),
\begin{equation}
\tau_b \equiv {1\over
F}{{k_B}\over{m_p}}\int_0^{z_b} dz\, \rho T .
\label{eq:definetaub}
\end{equation}
This time constant is closely related to the conventional thermal
relaxation time, the latter being defined as $\tau_{\rm th}\equiv {1/
F} (k_B/m_p) \int_0^{z_b} dz\, \rho T c_p$; in an isentropic,
ionised hydrogen plasma, $\tau_b \approx 2/5\, \tau_{\rm th}$.  

The first term on the right-hand side of equation
\refnew{eq:LtCv-dQall} quantifies the heat absorbed by the convective
region above $z_{b_0}$ when its entropy rises uniformly by $\Delta
s$. This entropy variation is constant throughout most of the region
because convection carries most of the stellar flux and because the
reaction of the convection towards pulsation is fast (Paper I, also
see Fig. \ref{fig:LtCv-sprofile}). The second term in equation
\refnew{eq:LtCv-dQall} represents the heat required in expanding or
evaporating the convection zone. This term is essential for
introducing nonlinearity into the flux variation.

Discarding terms higher than second order in $\Delta s$, we convert
equation \refnew{eq:firstlayprelim} into,
\begin{eqnarray} 
\left({{\delta F}\over F}\right)_{\rm ph}   
&\approx & \left({{\delta F}\over F}\right)_{b} - \tau_{b_0} {{d\Delta
s}\over{dt}} - (\tau_b(t) - \tau_{b_0}) {{d\Delta s}\over{dt}}\nonumber \\
& &- {{d z_b (t)}\over{dt}} \left[ \rho T {{k_B}\over{m_p}} (s -
s_0)\right]_{z_b(t)} \nonumber \\ &\approx &\left({{\delta F}\over
F}\right)_{\rm b} -
\tau_b (t) {{d\Delta s}\over{dt}}. \label{eq:LtCv-firstlaw}
\end{eqnarray}
We use $s-s_0 = 0 $ at $z_b (t)$ as the entropy perturbation in the
region $z > z_b(t)$ is assumed to be zero. Equation
\refnew{eq:LtCv-firstlaw} is similar to equations (40) and (42) in
Paper I, except that in the present case $\tau_b$ is time-dependent.

The entropy variation, $\Delta s$, is intimately related to the
photospheric flux perturbation (eqs. [38]-[39] in Paper I) as,
\begin{equation}
\Delta s = (B+C) \left({{\delta F}\over F}\right)_{\rm ph},
\label{eq:LtCv-sbdFph}
\end{equation}
where the dimensionless numbers $B$ and $C$ quantify respectively the
response of the photosphere and the superadiabatic region towards
gravity-mode pulsation (eqs. [25] \& [36] in Paper I). In ZZ Ceti
stars, $B$ and $C$ are both of order $8$.  Again, we take $(B+C)$ to
be $(B+C)(t)$. 

We define the following dimensionless derivatives,
\begin{eqnarray}
\beta & \equiv & {{\partial \ln (B+C)}\over{\partial \ln F}}
= {1\over 4} {{\partial \ln (B+C)}\over{\partial \ln T_{\rm
eff}}},  \nonumber \\
\gamma & \equiv & {{\partial \ln \tau_b}\over{\partial \ln F}}
= {1\over 4} {{\partial \ln \tau_b}\over{\partial \ln T_{\rm eff}}},
\label{eq:LtCv-bgnumber}
\end{eqnarray}
and combine equations \refnew{eq:LtCv-firstlaw},
\refnew{eq:LtCv-sbdFph} and \refnew{eq:LtCv-bgnumber} to produce
\begin{equation} 
\left({{\delta F}\over F}\right)_{\rm b} = X + \tau_{c_0} [1+(2 \beta + \gamma) X ] 
{{dX}\over{dt}}.
\label{eq:LtCv-Xeqn}
\end{equation}
Here, $\tau_{c} \equiv (B+C) \tau_{b}$, and $X$ is the simplification
of $ (\delta F/F)_{\rm ph}$. In the temperature range of ZZ Ceti
variables, our mixing length models yield $\beta \sim 1.2$ and $\gamma
\sim -15$ (obtained using Fig. 1 of Wu \& Goldreich [1999]). Note that
$2 \beta + \gamma < 0$; the thermal content of the convection zone
increases with decreasing effective temperature. The magnitude of the
nonlinear term $|(2\beta + \gamma) X|$ falls not far below unity when
$X$ is of order a few percent. Mathematically, equation
\refnew{eq:LtCv-Xeqn} also describes the motion of a pendulum with a
velocity-dependent mass under the influences of a periodic external
force and viscosity.

\subsection{Solutions
\label{subsec:LtCv-solu}}

We solve for $(\delta F/F)_{\rm ph}$ ($\equiv X$) using equation
\refnew{eq:LtCv-Xeqn}, first for the case when the input signal is 
a single mode and then for the case of two modes.

\subsubsection{{\it{Solution for a Single Mode}}
\label{subsubsec:LtCv-single}}

At any point on the stellar surface, for a single sinusoidal input of
the form
\begin{equation}
\left({{\delta F}\over F}\right)_{\rm b} = A_i \cos{( \omega_i t + \Psi_i)},
\label{eq:LtCv-input} \end{equation}
the solution for $(\delta F/F)_{\rm ph}$ can be expanded into
\begin{equation}
\left({{\delta F}\over F}\right)_{\rm ph} 
= a_i \cos{(\omega_i t + \psi_i)} + a_{2i} \cos{(2\omega_i t + \psi_{2i})}
+ ... , \label{eq:LtCv-testsingle}
\end{equation}
with the amplitudes and phases obtained from equation \refnew{eq:LtCv-Xeqn} as,
\begin{eqnarray}
a_i & = & {A_i \over{\sqrt{1 + (\omega_i \tau_{c_0})^2}}}, \nonumber \\
a_{2i} & = & {{a_i^2}\over{4}} {{|2\beta+\gamma| (2\omega_i
\tau_{c_0})}\over{\sqrt{1+ (2\omega_i \tau_{c_0})^2}}}, 
\label{eq:LtCv-value123}
\end{eqnarray}
and 
\begin{eqnarray}
\psi_i & = & \Psi_i - \arctan (\omega_i \tau_{c_0}), \nonumber \\ 
\psi_{2i}  & = & 2 \psi_i + \arctan\left({1\over{2\omega_i \tau_{c_0}}}\right) . 
\label{eq:LtCv-psi12}
\end{eqnarray}
The expressions for $a_i$ and $\psi_i$ are identical to equation (66)
in Paper I; the surface flux perturbation is reduced in magnitude and
delayed in phase relative to that entering the convection zone, as a
result of the heat retention (or release) of the convection zone
during pulsation. Typically in pulsating white dwarfs, $a_i$ ranges
from a few tenths of a percent to a few percent. The fact that the
harmonic frequency of a such a weak mode is actually observable is
largely because $|2 \beta + \gamma| \gg 1$. The amplitude of the
second harmonic ($a_{3i}$), not considered here, is of order $|
\beta \gamma| a_i^3$, and should be detectable when $a_i$ is large.

The above solution is best understood in terms of light curve
distortion. The phase lag and the reduction factor between $(\delta
F/F)_{\rm ph}$ and $(\delta F/F)_b$ scale with the thickness of the
convection zone. Since the convection zone is at its thinnest when
$(\delta F/F)_{\rm ph}$ approaches its maximum, one expects the phase
lag and the reduction factor to be smaller at the maximum of $(\delta
F/F)_{\rm ph}$ than at the minimum. This leads to peaked light curves
with sharp ascent and shallow descent. Fourier transform of such a
light curve exhibit harmonics that lead the fundamentals in time,
$\psi_{2i} - 2 \psi_i > 0$ \cite{brick92}.

\subsubsection{{\it{Solution for Two Modes}}
\label{subsubsec:LtCv-two}}

When $(\delta F/F)_b$ is comprised of two sinusoidal signals with
radian frequencies $\omega_i$ and $\omega_j$,
\begin{equation}
\left({{\delta F}\over F}\right)_{\rm b} = A_i \cos{(\omega_i t + \Psi_i)} + 
A_j \cos{(\omega_j t + \Psi_j)},
\label{eq:LtCv-input2} \end{equation}
the following form of solution is adopted,
\begin{eqnarray}
\left({{\delta F}\over F}\right)_{\rm ph} 
& = & a_i \cos{(\omega_i t + \psi_i)} + a_{2i} \cos{(2\omega_i t +
\psi_{2i})}   \nonumber \\
& & + a_j \cos{(\omega_j t + \psi_j)} + a_{2j}
\cos{(2\omega_j t + \psi_{2j})} 
\nonumber \\ 
& & + a_{i-j}
\cos{((\omega_i - \omega_j) t + \psi_{i-j})} \nonumber \\
& & +  a_{i+j} \cos{((\omega_i + \omega_j) t + \psi_{i+j})} + ... \, .
\label{eq:LtCv-testtwo}
\end{eqnarray}
Subscripts for the different coefficients are chosen to represent
their corresponding frequencies. We obtain the following generalized
expressions for the amplitudes and phases of the combination
frequencies (including both harmonics and mixed combinations),
\begin{eqnarray}
a_{i\pm j} & = & {n_{ij}\over 2}{{a_i a_j}\over 2}{{|2 \beta + \gamma|
(\omega_i \pm \omega_j) \tau_{c_0}}\over{\sqrt{1+ ((\omega_i \pm
\omega_j) \tau_{c_0})^2}}}, 
\label{eq:general_combin}\\
\psi_{i\pm j} & = & (\psi_i \pm \psi_j) + 
\arctan\left({1\over{(\omega_i \pm \omega_j) \tau_{c_0}}}\right).
\label{eq:general-phase}
\end{eqnarray}
where $n_{ij}$ counts the number of possible permutations given $i$
and $j$: $n_{ij} = 2$ if $i \neq j$, and $1$ if otherwise. 

\subsection{Angular Integration and Bolometric Corrections
\label{subsec:angularbolo}}

Equations \refnew{eq:general_combin}-\refnew{eq:general-phase}
quantify the amplitudes and the phases for a combination frequency
at every point on the stellar surface. In this section, we relate
them to the observable quantities.

Defining $\Theta$ and $\Phi$ to be the spherical coordinates in the
stellar rotating frame, we adopt the following form for the flux
perturbation incident upon the convective bottom,
\begin{equation}
\left( {{\delta F}\over F}\right)_b = \sum_i A_i P_{\ell_i}^{m_i}(\Theta, \Phi) 
\cos(\omega_i t + m_i \Phi+ \Psi_{i_0}).
\label{eq:total_sum} 
\end{equation}
Here, $P_{\ell}^m e^{i m \Phi} = Y_{\ell m} $, the latter being the
spherical harmonic function normalized to unity over the sphere. The
$\Psi_i$ appearing in equation \refnew{eq:LtCv-input} is now $m_i \Phi
+ \Psi_{i_0}$, while $\Psi_{i_0}$ is a constant over angle and time.
Similarly, the linear part of the photospheric flux variation
can be written as
\begin{equation}
\left( {{\delta F}\over F}\right)_{\rm ph} = \sum_i a_i P_{\ell_i}^{m_i}(\Theta, \Phi) 
\cos(\omega_i t + m_i \Phi+ \psi_{i_0}),
\label{eq:sum_dFFph} 
\end{equation}
with $a_i$ and $\psi_{i_0}$ related to $A_i$ and $\Psi_{i_0}$ as in equations
\refnew{eq:LtCv-value123} - \refnew{eq:LtCv-psi12}.

An observer with a line-of-sight inclined relative to the rotation
axis by an angle $\Theta_0$\footnote{Degeneracy in $\Phi$ implies that
we can take $\Phi_0 = 0$.} would detect the following bolometric flux
variation 
\begin{equation} 
\left( {{\delta f}\over f}\right)_{\rm bol}= \sum_i a_i
g_{\ell_i}^{m_i} (\Theta_0) \cos(\omega_i t + \psi_{i_0}),
\label{eq:sum_dLL}
\end{equation}
where the factor $g_{\ell}^m$ includes effects such as geometric
projection and limb-darkening. In Table \ref{table:appendix-gvalue},
we present the expression for $g$ as well as its values when $\ell \leq
2$.

The angular dependence of a combination frequency is described by the
product of the angular dependences of its principal modes.  This
arises because equation \refnew{eq:general_combin} is valid for every
point on the stellar surface. We express the corresponding
disc-integrated flux variations as
\be
\left( {{\delta f}\over f}\right)_C=  \sum_{i,j} a_{i\pm j}
G_{\ell_i \,\, \ell_j}^{m_i\pm m_j} (\Theta_0)
\cos((\omega_i\pm \omega_j) t + \psi_{i_0 \pm j_0}).
\label{eq:sum_dLL_C}
\ee
The expression for the $G$ function is presented in the appendix,
together with some useful values of $G$.

Broad-band photometric observations produce amplitude variations that
are related to the bolometric variations as $(\delta f/f) =
\alpha_\lambda (\delta f/f)_{\rm bol}$.  For ZZ Ceti stars observed in
V-band, $\alpha_V \sim 0.4$.

The theoretical expectation value for $R_c$, the ratio between the
observed amplitude of a combination and the observed amplitudes of its
principal modes (van Kerkwijk \etal 1999), is therefore,
\begin{eqnarray}
R_c & \equiv & {
{\left({{\delta f}\over f}\right)_{i\pm j}}
\over
{n_{ij}
{\left({{\delta f}\over f}\right)_{i}}
{\left({{\delta f}\over f}\right)_{\rm j}}}} \nonumber \\
& = &
{ {|2 \beta + \gamma| (\omega_i \pm \omega_j) \tau_{c_0}}
\over{ 4 \alpha_V \sqrt{1+ ((\omega_i \pm \omega_j) \tau_{c_0})^2}}} 
{{ G_{\ell_i \,\, \ell_j}^{m_i\pm m_j}}\over{g_{\ell_i}^{m_i} g_{\ell_j}^{m_j}}}.
\label{eq:theory-Rc}
\end{eqnarray}

\section{What can be Learned?}
\label{sec:learn}

In this section, we discuss the potential of extracting information
from measurements of the combination frequencies. We first describe
what information may be available, and then compare the observations
of two variable white dwarfs (a DA and a DB) with our analytical
results.

\subsection{Prelude}
\label{subsec:prelude}

\begin{figure}
\centerline{\psfig{figure=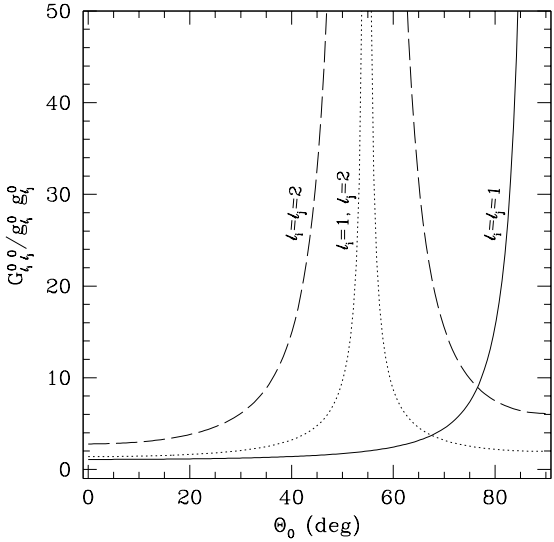,angle=0,width=8.5cm}}
\caption[]{The ratio $G_{\ell_i \ell_j}^{0 \,\, 0}/(g_{\ell_i}^0 
g_{\ell_j}^0)$ as a function of the inclination angle $\Theta_0$ when
$\ell_i$, $\ell_j$ take $1$ or $2$. Notice that at certain angles, the
combination frequencies may be observable while unfavourable
geometrical cancellation renders their principal modes invisible. }
\label{fig:LtCv-Gfunctions}
\end{figure}

A major difficulty of white dwarf asteroseimology lies in our
inability to securely identify the spherical degree ($\ell$) for the
pulsation modes. In the cases where combination frequencies are
detected, one could use the observed values of $R_c$
(eq. [\ref{eq:theory-Rc}]) to determine the $\ell$ value for the
principal modes. To illustrate this possibility, we consider the
harmonic of a $m = 0$ principal mode. The ratio $G_{\ell \ell}^{0
0}/(g_{\ell}^0 g_{\ell}^0)$ (and consequently the value of $R_c$) is
significantly higher for $\ell= 2$ than for $\ell=1$, except when
$\Theta_0 $ approaches $90^{\circ}$ (see
Fig. \ref{fig:LtCv-Gfunctions} and eq. [\ref{eq:Gvalueforl}]). This
arises as the apparent amplitudes of higher $\ell$ modes suffer
stronger cancellation when integrated over the stellar disc while the
harmonics of these modes do not. Note that this is purely a geometric
argument and a similar method of $\ell$ identification would work not
only for the sum or difference combinations of two principal modes in
white dwarfs, but also for other variable stars that exhibit
combination frequencies and where the amplitude of a combination
frequency satisfies $a_{i\pm j} \propto a_i a_j$ (as in
eq. [\ref{eq:general_combin}]) for every point on the stellar surface.

Inside the pulsational instability strip, the thermal time constant of
the convection zone ($\tau_{c_0}$) varies monotonically and
sensitively with the stellar effective temperature. The value of
$\tau_{c_0}$ discloses the relative location of a variable in the
strip. In addition, we can study convection under the white dwarf
environment if we can empirically determine the $T_{\rm
eff}$-$\tau_{c_0}$ relation. Time-resolved spectroscopy provides one
way to measure $\tau_{c_0}$ \cite{ltcv-marten98}. However, this
technique requires large telescopes and works only for relatively
bright white dwarfs. What about using the combination frequencies?

The relative phase between a combination frequency and its principal
modes ($\psi_{i_0 \pm j_0} - (\psi_{i_0} \pm \psi_{j_0})$) yields
$\tau_{c_0}$ straightforwardly (eq. [\ref{eq:general-phase}]). This
phase depends on $\tau_{c_0}$ more sensitively at low
frequency. However, care needs to be taken to avoid systematic effects
that affect phase measurements adversely, such as the presence of
small neighbouring periodicities that are not accounted for.

Another way to measure $\tau_{c_0}$ is to take the ratio between the
amplitudes of the sum and the difference combinations from the same
pair of principal modes, (eq. [\ref{eq:general_combin}]) the geometric
factor cancels when one or both of $m_i$, $m_j$ is $0$.
\begin{equation}
{{\left({{\delta f}\over f}\right)_{i+j}}\over {\left({{\delta
f}\over f}\right)_{i-j}}} = {{(\omega_i + \omega_j)}\over{(\omega_i -
\omega_j)}} {{\sqrt{1+ (\omega_i - \omega_j)^2 \tau_{c_0}^2}}\over
{\sqrt{1+ (\omega_i + \omega_j)^2 \tau_{c_0}^2}}} {{G_{\ell_i \,\,
\ell_j}^{m_i + m_j}}\over {G_{\ell_i\,\, \ell_j}^{m_i - m_j}}}.
\label{eq:LtCv-c2d2ratio}
\end{equation}
Note that amplitudes measured in the lower frequency region are
generally less accurate due to higher noise levels.

The two dimensionless numbers, $\beta$ and $\gamma$, quantify the
deepening of the convection zone when a white dwarf cools. It is
therefore related to the width of the white dwarf instability strip.
Let us associate the blue edge of the ZZ Ceti instability strip
($T_{\rm eff} \approx 12,000 \K$) with $\tau_{c_0} = 20 \s$ (when the
lowest order $\ell = 1$ gravity-mode mode satisfies $\omega \tau_{c_0}
= 1$, see Paper I), and the red edge of the strip with $\tau_{c_0} =
1300 \s$ (when the $1000 \s$ period mode becomes invisible at the
surface, $\omega \tau_{c_0} = 10 \gg 1$, see Paper I). We find the
width of the instability strip to be $\sim 1000 \K$ when we adopt
$\beta + \gamma \sim -14$ as in \S \ref{sec:LtCv-origin}. A larger
value of $|\beta + \gamma|$ would correspond to a narrower instability
strip. These numbers can be obtained from combination frequency
measurements together with other unknown quantities.

A number of practical difficulties may arise in the actual analysis.
For instance, different $m$ components of a gravity-mode are closely
spaced in frequency and may not be resolved by observations of short
duration, whereas in observations of sufficiently long duration
temporal changes in the amplitudes of pulsation may occur. In the
following sections, we apply our results ignoring these difficulties.

\subsection{The DB variable GD358\label{subsec:LtCv-gd358}}

For our analysis of the DB variable GD358, we use the Whole Earth
Telescope (WET) data \cite{winget94} in which different $m$ components
of the principal modes are well resolved. We assume that mode
amplitudes do not change appreciably during the entire observation.

\begin{table*}
\begin{center}
\begin{tabular}{ccccccccc}
\hline
\multicolumn{4}{c} {principal modes} & \mbox{   } & \multicolumn{4}{c} {combination  frequencies}\\
\multicolumn{2}{c} {$k = 13$} & \multicolumn{2}{c} {$k=15$}  
&  \mbox{   } & $\nu$ &\multicolumn{2}{c} 
{observed amplitudes} & {theoretical} \\
$m_i$ & (mma) & $m_j$ & (mma) & \mbox{   } & ($\mu$Hz) & (mma) & relative &
 relative \\
\hline
        & & & & & & & & \\ 
-1 & 6.28 & -1 & 9.33 &  \mbox{   } & 3033.07 & $<0.5$ & $< 0.23$ & 0.23 \\
 0 & 5.78   & -1 & 9.33 & \mbox{   } & 3038.65 & $<1.0$ & $< 0.5$ & 0.21\\
-1 & 6.28 & 0 & 19.03 &  \mbox{   } & 3039.07 &  1.55 & 0.72 & 0.46\\
 0 & 5.78 & 0 & 19.03 &  \mbox{   } & 3044.65 & 2.14 & 1.00 &  1.00 \\
 +1 & 5.46  & -1 & 9.33 &  \mbox{   } & 3044.76 & $<1.5$ & $< 0.7$ &  0.73\\
-1 & 6.28    & +1 & 9.27 &  \mbox{   } & 3045.84 &  1.79 & 0.84 &   0.84\\
 +1 & 5.46    & 0 & 19.03 & \mbox{   } & 3050.76 &1.10 & 0.50 &  0.40 \\
 0 & 5.78     & +1 & 9.27 & \mbox{   } & 3051.42 & 0.77 & 0.36 &  0.21\\
+1 & 5.46    & +1 & 9.27 & \mbox{   } & 3057.53 & 0.92 & 0.43 &  0.19 \\
       &    &         &    &      &  &    &    &   \\
\hline
\end{tabular}
\end{center}
\caption[]{Pulsation amplitudes for different components in the
$(k=13) + (k=15)$ combination. The observed values are extracted from
Figure 8 of Winget \etal \shortcite{winget94}, where the unit $mma$
stands for milli-modulation amplitude and equals to $0.1 \%$. The
amplitude of the $(m_i = -1) + (m_j = -1)$ component falls below
detection limit. The $(m_i = 0 ) + (m_j = -1)$ and $(m_i = +1) + (m_j
= -1)$ components are too close to other components to be resolved. We
give rough upper limits to their amplitudes. All relative amplitudes
are amplitudes scaled by the $(m_i = 0) + (m_j=0)$ component at $\nu =
3044.5 \mu$Hz. Here, $\nu$ is the angular frequency and is related to
$\omega$ by $\omega = 2 \pi \nu$. The theoretical relative amplitudes
are calculated using equation \refnew{eq:theory-Rc} with $\Theta_0 =
45^{\circ}$. We find that the data are roughly consistent with
inclination angles between $40^{\circ}$ and $50^{\circ}$.}
\label{tab:LtCv-multiplet}
\end{table*}

\subsubsection{{\it{Fine Structure in the Combination Frequencies}}
\label{subsubsec:LtCv-fine}}

The sum (or difference) combination of two $\ell = 1$ principal modes
(each split into three $m$ components) can contain up to nine
fine-splitting components that are different in frequency
\cite{winget94}. The relative strengths among these components reflect
their respective projection in the direction of our line-of-sight. In
Table \ref{tab:LtCv-multiplet}, we compare the observed and the
theoretically expected values of these relative strengths for the sum
combination of two principal modes, $k= 13$ and $k=15$ (see Fig. 8 of
Winget \etal 1994). We find overall agreements when the inclination
angle $\Theta_0$ falls within the range between $40^{\circ}$ and
$50^{\circ}$.  Some mismatches exist. They may be caused by the fact
that some combination components are too close in frequency to be
resolved by the WET run. In addition, the amplitudes of the principal
modes vary during the run and this may affect the comparison.  We notice
that the WET run lasted $11$ days which is of order the natural growth
time for the $k=13$ and $k=15$ modes \cite{paperII}.

\subsubsection{ {\it{Combination Frequencies at Large}}
\label{subsubsec:LtCv-gd358combi}}

\begin{figure}
\centerline{\psfig{figure=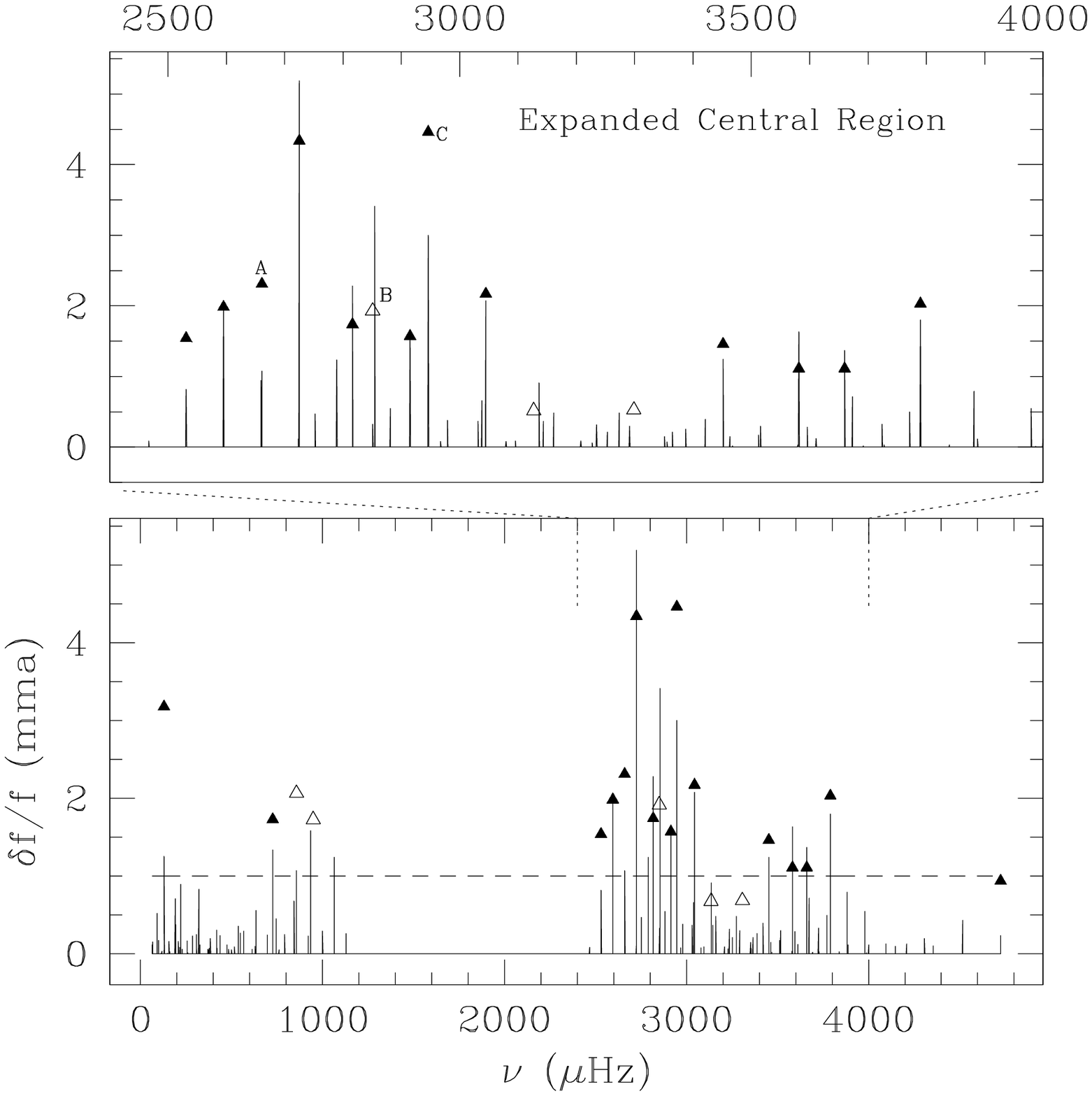,angle=0,width=10.0cm}}
\caption[]{Comparison between observed (triangles) and theoretical
amplitudes (vertical lines) for the combination frequencies in GD
358. We calculate all possible two-mode combinations for the $m=0$
gravity-modes listed in Table 2 of Winget \etal (1994). A triangle is
solid if the observed frequency of the combination lies within $1
\mu$Hz of its expected value, where $1 \mu$Hz is roughly the frequency
resolution for this WET run. With this we test the assumption that a
combination is produced by $m=0$ principal modes. The top panel
expands the central region of the lower panel, while the horizontal
dashed line in the lower panel is the detection limit in this WET
observation.  Three letters (`A',`B' and `C') denote three cases of
serious disagreement.}
\label{fig:LtCv-GD358b}
\end{figure}

Winget \etal (1994, Table 3) determined pulsation power for the
strongest component in each combination. We compare their data with
theoretical estimates in Figure \ref{fig:LtCv-GD358b}. For our
analysis, we assume that all principal modes have spherical degree
$\ell = 1$ and that the strongest component in each combination is
produced by the $m = 0$ components in the principal modes. The second
assumption is tested in Figure \ref{fig:LtCv-GD358b}. As in \S
\ref{subsubsec:LtCv-fine}, We adopt $\Theta_0 = 45^{\circ}$. We
further take $|2 \beta + \gamma|/\alpha_V = 25$ and $\tau_{c_0} = 300
\s$ to produce theoretical estimates. Such a choice of $\tau_{c_0}$
ensures that all gravity-modes observed in GD358 satisfy the necessary
condition for overstability, $\omega \tau_{c_0} > 1$ \cite{paperI}.
Varying $\tau_{c_0}$ from $10 \s$ to $1000 \s$ only changes the
estimates by $\sim 40\%$.

Theoretical estimates based on the above choices of parameters largely
reproduce the observed values. Figure \ref{fig:LtCv-GD358b} shows that
most ($\sim 90\%$) combinations that are expected to have amplitudes
above the $\sim 1$ mma detection limit are indeed observed.  However,
a few significant discrepancies are present and they merit some
discussions.

The power at $\nu = 2660.84 \mu$Hz (labelled with `A' in the upper
panel of Fig. \ref{fig:LtCv-GD358b}) is attributed to the
$(k=18)+(k=15)$ combination by Winget \etal (1994). The observed
amplitude is roughly twice the theoretical one. Interestingly enough,
the $(k=16)+(k=17)$ combination lies at $\nu = 2659.43 \mu$ Hz and is
expected to reach approximately the same amplitude as the former
combination. This may explain the mismatch seen at `A' if the two
combinations are not well resolved from each other.

The combination at $2848.28 \mu$Hz (`B') lies $\sim 6 \mu$Hz away from
the harmonic of the $k=15, m=0$ mode and is possibly the sum
combination between this mode and the $k=15, m=-2$ mode.  As the
$k=15, m=0$ mode has the greatest amplitude in the Fourier spectrum,
it is surprising that we do not observe its harmonic.\footnote{Even
though $k=15$ is the strongest mode, its harmonic is expected to have
lower amplitude than, for instance, the combination $(k=15)+(k=17)$
due to the factor $n_{ij}$ in equation \refnew{eq:general_combin}.}
This problem is not unique to GD 358, as our investigation of ZZ Psc
(\S \ref{subsec:LtCv-ZZPsc}) finds. Full numerical simulations of the
convective response may provide a solution to this problem (Ising \&
Koester 1999).

We can not explain the disagreement seen at $2946.65 \mu$Hz (case
`C') either.

Combinations at low frequencies are potentially most rewarding for
estimating $\tau_{c_0}$. However, the signal-to-noise ratio is lower at
these frequencies.

\subsection{The DA variable ZZ Psc \label{subsec:LtCv-ZZPsc}}

ZZ Psc (aka G29-38) is a relatively cool DA variable exhibiting large
pulsation amplitudes that vary in time \cite{kleinman95}. We
investigate the combination frequencies in this star using a set of
time-resolved spectroscopy data taken with the Keck II telescope
\cite{ltcv-marten98}.  The observation lasted five hours and could not
resolve the rotational splitting.  However, the signal-to-noise ratio
is high and phases of pulsation are measured from the data. These can
be compared with equation \refnew{eq:general-phase}.

Among the handful of eigenmodes seen in the spectra, one (marked as `F4')
stands out as an $\ell = 2$ mode (Clemens \etal 1999), while all
others have $\ell = 1$. This result is supported by the fact that `F4'
is associated with a relatively large line-of-sight velocity (van
Kerkwijk \etal 1999). In this section, we show that the combination
frequencies also provide confirmation for this $\ell$-identification.

\begin{figure}
\centerline{\psfig{figure=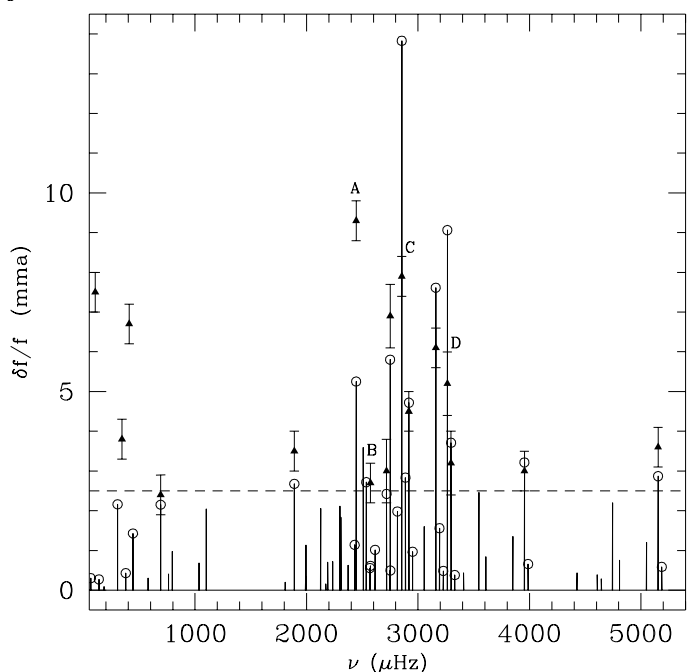,angle=0,width=10.0cm}}
\caption[]{Amplitudes of combination frequencies in ZZ Psc. The
observed values are plotted as solid triangles with error bars. The
vertical lines are theoretical amplitudes of all possible combinations
between observed gravity-modes. They are crowned with a open circle if
they lie within $50 \mu$Hz in frequency from the observed
combinations. This is roughly the frequency resolution for this set of
data. The noise level in the data is indicated by the horizontal
dashed line. }
\label{fig:LtCv-ZZPscd}
\end{figure}

We first assume all observed principal modes are $\ell = 1$,
$m=0$. Taking $\tau_{c_0} = 250\s$,\footnote{This choice of
$\tau_{c_0}$ ensures that all observed g-modes satisfy $\omega
\tau_{c_0} > 1$, and it puts ZZ Psc close to the red edge of the DA
instability strip.}  $|2 \beta + \gamma| = 10$ and $\Theta_0 =
30^{\circ}$,\footnote{In this case, we can not constrain $|2\beta
+\gamma|$ and $\Theta_0$ separately.} we produce a theoretical
amplitude spectrum for the combination frequencies
(Fig. \ref{fig:LtCv-ZZPscd}), onto which we plot the observed
amplitudes as well as their error bars. The comparison is
satisfactory; all combinations that are estimated to lie above the
noise level are indeed detected. However, it is a surprise to find the
three lowest frequency combinations to have much higher amplitudes
than expected. As signals at very low frequencies may suffer from
larger noise, this discrepancy needs to be confirmed by future
observations.

In Figure \ref{fig:LtCv-ZZPscd}, the combinations noted by `A' and `D'
are respectively the first harmonics of the second largest and the
largest g-modes. Equation \refnew{eq:general_combin} under-predicts
the amplitude for `A', and over-predicts it for `D'. The latter
problem, interestingly enough, appears in GD358 as well (see
Fig. \ref{fig:LtCv-GD358b}). This may indicate that our perturbation
analysis fails at large mode amplitudes. The combination at `C' (sum
of the strongest and the second strongest modes) falling much below
the theoretical estimate jibes with this suggestion.

\begin{figure}
\centerline{\psfig{figure=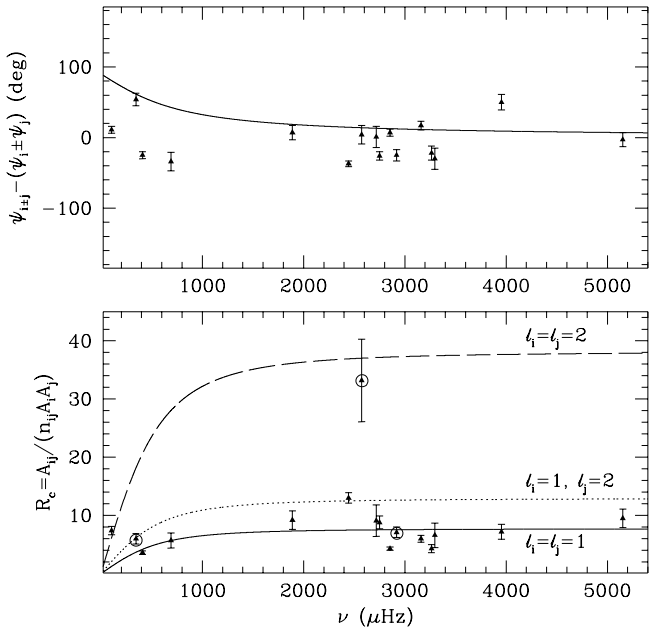,angle=0,width=9.5cm}}
\caption[]{For ZZ Psc, we compare observed values of relative phase
and amplitude ratio between combination frequencies and their
principal modes, plotted as triangles, with theoretical expectations.
Relative phase (eq. [\ref{eq:general-phase}]) does not depend on the
spherical harmonic degrees of the two principal modes, while amplitude
ratio ($R_c$ in eq. [\ref{eq:theory-Rc}]) does. We plot $R_c$ for
$\ell = 1$ and $2$ principal modes. The combination with the largest
$R_c$ value is the harmonic of mode `F4', while the other two circled
combinations are mixing results of `F4' with other principal modes.}
\label{fig:LtCv-Rcpsi}
\end{figure}

Clemens \etal (1999) and van Kerkwijk \etal (1999) argued that mode
`F4' has spherical degree $\ell = 2$. If this is true, at $\Theta_0 =
30^{\circ}$, its harmonic would have an amplitude $6$ times larger than
if it was $\ell = 1$ (see \S \ref{subsec:prelude} and
Fig. \ref{fig:LtCv-Gfunctions}).  This could explain the disagreement
at `B'. In the lower panel of Figure \ref{fig:LtCv-Rcpsi}, we show
theoretical $R_c$ values for combinations of $\ell = 1$ and $2$
principal modes. The observed $R_c$ value for `B' is consistent with
it being the harmonic of an $\ell = 2$ mode. This is a piece of
independent evidence supporting the $\ell = 2$ identification for mode
`F4'. At $\Theta_0 = 30^{\circ}$, $R_c$ values for combinations that
involve `F4' and another mode do not differ appreciably from other
combinations (see Fig. \ref{fig:LtCv-Gfunctions}).

In the upper panel of Figure \ref{fig:LtCv-Rcpsi}, we study the phase
difference between a combination and its principal modes. From
equation \refnew{eq:general-phase}, one expects most of these phase
differences to be positive and to lie close to $0$. Indeed, the
observed phase differences cluster closely around $0$. However, in
detail the fit is not good. This may be related to the unaccounted-for
periodicities which influence the phase determinations (van Kerkwijk
\etal 1999)

Employing data from WET and other observations (see Kleinman \etal
1998), Vuille \shortcite{vuille00b} analyzed the relative phases of
combination frequencies in ZZ Psc.  He found $\psi_{i\pm j} - (\psi_i
\pm \psi_j)$ to be small but predominantly positive. This reflects the
fact that pulsation light-curves have sharper ascent than
descent. Vuille concluded that these relative phases remain fairly
constant over the years, a finding consistent with equation
\refnew{eq:general-phase}.

\section{Conclusions}
\label{sec:conclusions}

\subsection{DA \& DB Variables
\label{subsec:DADB}}

Our analysis leads to two key formulae (eq.  [\ref{eq:theory-Rc}] and
[\ref{eq:general-phase}]) that describes the strength and the phase of
the combination frequencies relative to their principal modes.

A few stellar parameters enter these formulae. They are the thermal
constant of the stellar convection zone at equilibrium ($\tau_{c_0}$),
the rate of deepening of the convection zone with cooling of the star
(quantified by $|2 \beta + \gamma|$), and the inclination angle
between the observer's line of sight and the stellar pulsation axis
($\Theta_0$).  It is becoming possible to use the combination
frequencies to constrain these stellar parameters. We find that for
both GD358 and ZZ Psc, the observed amplitude spectra can be roughly
reproduced using reasonable choices of the above parameters.  The same
choices can also explain the values for the dimensionless numbers
($a,b$ and $c$) Brickhill (1992) summarized from his numerical
study. The $\ell$ and $m$ values of the eigenmodes also enter into
these formulae.  This presents the potential of determining the $\ell$
values of the principal modes using the combination frequencies. An
$\ell = 2$ mode is expected to have a stronger harmonic than an $\ell
= 1$ mode, and this is indeed observed in ZZ Psc.

When analyzing observed combination frequencies, we have ignored
amplitude variability during a long observing run, or in the case of a
short run, have assumed that modes are axisymmetric with respect to
the pulsation axis ($m=0$).  The failure of these assumptions may
account for some of the discrepancies between observation and theory
and may prevent us from accurately determining stellar parameters. More
suitable data sets might yield more conclusive information.

We find that theory over-predicts the amplitude in the harmonic of the
strongest pulsation mode in the two stars we considered.  We suspect
that it results from the stronger nonlinearity associated with the
largest mode. 

Combination frequencies are produced by the surface convection zone in a
pulsating white dwarf. Photosphere in these stars is not thermally
capable of distorting the light curve. We therefore expect equations
\refnew{eq:theory-Rc} and \refnew{eq:general-phase} to hold for all
wavelengths. 

\subsection{Other Types of Variables
\label{subsec:othertype}}

Combination frequencies have also been reported in two PG1159
variables (PG1707+427, Fontaine \etal 1991; HS2324+3944, Silvotti
\etal 1999). Presumably, these hot white dwarfs do not have surface
convection zones. What could be distorting the light curves?

Could a radiative, partially ionising layer produce the distortions?
Such a layer is believed to exist in the upper atmosphere of PG1159
variables and is believed to be responsible for driving the observed
pulsations. It is similar to the surface convection zones in DA and DB
variables in that it retains heat when warmer, and releases heat when
cooler. However, unlike in the case of the convection zones, the
amount of heat retained (or released) by the partial ionising region
can not be significantly modulated throughout the pulsation cycle by
the presence of other pulsation modes. This is because the reaction
time of the ionising region is roughly the local thermal relaxation
time, which is of the same order as periods of overstable modes. Thus,
we can not explain the combination frequencies in PG 1159 stars.

We note that in other types of small amplitude pulsators, e.g.,
$\delta$-Scuti stars, sdB variables, and $\gamma$-Doradus stars,
combination frequencies have also been reported. We conjecture that a
thin surface convection zone is present in these variables and is
capable of exciting pulsation modes, as well as distorting the light
curves.


\bigskip

The author would like to acknowledge the many beneficial comments and
suggestions by Drs. Peter Goldreich, Joerg Ising, Marten van
Kerkwijk, Scot Kleinman and Francois Vuille.


\begin{table*}
\begin{center}
\begin{tabular}{c|ccc|}
$\ell \setminus |m|$& $2$ & $1$ & $0$ \\
\hline
$0$	& -- & -- & 0.28 \\
$1$	& -- & $0.245\sin\Theta_0$ & $0.346\cos\Theta_0$ \\
$2$	&$0.126\sin^2\Theta_0$ & $0.126\sin(2\Theta_0)$ &$0.05[1+3\cos(2\Theta_0)]$ \\
\hline
\end{tabular}
\end{center}
\caption[]{Values of $g_{\ell}^m$ as functions of $\Theta_0$,
the inclination angle between the observer's line-of-sight and the
stellar rotation axis.}
\label{table:appendix-gvalue}
\end{table*}

\begin{table*}
\begin{center}
\begin{tabular}{c|ccc|}
$m_j \setminus m_i $ & $-1$ & $0$ & $+1$ \\
\hline
$-1$ & $0.65$ &
$0.65$ & $0.63+{{0.90}\over{\sin^2 \Theta_0}}$ \\
$0$ & $0.65$ & $0.65 + {{0.45}\over{\cos^2 \Theta_0}}$ &
$0.65$\\
$+1$ & $0.63 + {{0.90}\over{\sin^2 \Theta_0}}$ & $0.65$ & $0.65$
\\
\hline
\end{tabular}
\end{center}
\caption[]{Values of $G_{1\,\,\,\,\, 1}^{m_i + m_j}/(g_{1}^{m_i} g_{1}^{m_j})$ 
as functions of $\Theta_0$. Values of $G_{\ell_i\,\,\,\, \ell_j}^{m_i -
m_j}/(g_{\ell_i}^{m_i} g_{\ell_j}^{m_j})$ are obtained by reversing the
sign of $m_j$.}
\label{table:appendix-bigG-plus}
\end{table*}

\begin{appendix}

\section{Angular Integration
\label{subsec:LtCv-Ylm}}

Let $(\Theta, \Phi)$ be the spherical coordinate system defined by the
stellar rotation axis, and $(\theta, \phi)$ be that defined by the
observer's line-of-sight. Let $\Theta_0$ be the angle between this
line-of-sight and the rotation axis. The two coordinate systems are
related by
\begin{eqnarray}
\cos \Theta & = & - \sin \Theta_0 \sin \theta \cos\phi + \cos \Theta_0 \cos\theta,
\nonumber \\
\sin\Theta \cos\Phi & = & \cos \Theta_0 \sin\theta \cos\phi + \sin\Theta_0 \cos\theta,
\nonumber \\
\sin\Theta \sin\Phi & = & \sin\theta \sin\phi.
\label{eq:appendix-transform}
\end{eqnarray}

To obtain the photometric pulsation amplitude, we integrate the
photospheric flux variation over the visible disc,
\be
\left({{\delta f}\over f}\right)
 = {1\over{2 \pi}} {\oint_0^{2 \pi}\, d\phi\, \int_0^1\, h(\mu) \mu\,
 d\mu
\left({{\delta F}\over F}\right) (\Theta,\Phi,t)} , 
\label{eq:appendix-dLLtodFF}
\ee
where $\mu =\cos\theta$, and $h(\mu)$ is the limb-darkening function
normalized by $\int_0^1\, h(\mu)\mu\, d\mu = 1$. For our exercise, we
adopt the Eddington limb-darkening law of $h(\mu) = 1 + 3/2 \mu$ (see,
e.g., Shu 1991), as is appropriate for grey-atmosphere.

It is convenient to define
\be
g_{\ell}^m (\Theta_0) \equiv {1\over {2 \pi}}\oint_0^{2 \pi}\, d\phi
\int_0^1 h(\mu)\mu d\mu {\rm Re}[Y_{\ell}^m (\Theta, \Phi)].
\label{eq:appendix-gdefine}
\ee
We list values of $g$ in Table \ref{table:appendix-gvalue} for $\ell
\leq 2$.

The angular dependence of a combination frequency is described by the
product of the angular dependences of its principal modes. Its
photometric amplitude is related to its photospheric amplitude by the
following function,
\begin{eqnarray}
G_{\ell_i\,\, \ell_j}^{m_i\pm m_j} \equiv & &  {1\over {2 \pi}}\oint_0^{2 \pi}\,
d\phi \int_0^1 h(\mu) \mu d\mu \nonumber \\
 & & \hskip-1.0cm\times  P_{\ell_i}^{m_i}(\Theta,\Phi)  P_{\ell_j}^{m_j}(\Theta, \Phi) 
\cos((m_i \pm m_j) \Phi).
\label{eq:appendix-bigGdefine}
\end{eqnarray}
It is easiest to evaluate the $G$ function by reducing the above
product of the Legendre functions into a linear sum of such functions. We
present values of the ratio $G_{\ell_i\,\,
\ell_j}^{m_i + m_j}/(g_{\ell_i}^{m_i} g_{\ell_j}^{m_j})$ in Table
\ref{table:appendix-bigG-plus} 
for $\ell_i = \ell_j = 1$. Values of $G_{\ell_i\,\, \ell_j}^{m_i -
m_j}/(g_{\ell_i}^{m_i} g_{\ell_j}^{m_j})$ can be trivially obtained
from the same table by changing the sign of $m_j$.
For our discussion in \S \ref{subsec:prelude}, we also need
\begin{eqnarray}
{{G_{1\,\, 2}^{0\pm 0}}\over{g_1^0 g_2^0}} & = & {{1.97 + 0.82
\cos^2\Theta_0}\over{3 \cos^2\Theta_0 - 1}}, \nonumber \\
{{G_{2\,\, 2}^{0\pm 0}}\over{g_2^0 g_2^0}} & = & {{8.39 + 2.51 \cos(2
\Theta_0) + 0.22 \cos(4 \Theta_0)}\over{(3 \cos^2\Theta_0 - 1)^2}}.
\label{eq:Gvalueforl}
\end{eqnarray}

\end{appendix}

\end{document}